\documentstyle[12pt]{article}
\begin{document}

\title{Influence of the second viscosity on the flow propagation}
\author{V.V. Maximov
N.E. Molevich, I.P. Zavershinsky  and  A.P. Zubarev \\
\rm Samara State Aerospace University,\\ Moscow Shosse 34, Samara, 443086, Russia}
\date {$~$}
\maketitle
\begin{abstract}
We found the  energetic stability criteria of
non-equilibrium gase plain flows.
\end{abstract}
\vspace{10mm}

The second (volume) viscosity coefficient $\zeta$ can be negative in
nonequilibrium media. Medium with the negative second viscosity is acoustically
unstable. The nonlinear structures in the acoustically active gases is strongly
different from that in the equilibrium media. In present paper we investigate
some problems, connected with the second viscosity influence on flow
characteristics.

Let us consider the thin body ($l_x /l_y \gg M$), $M>1$
is the Mach number in the supersonic laminated stream of nonequilibrium gas.
It is proposed that the angle of attack $\delta=l_y /l_x \ll1$.
It follows in the standard way (\cite{6131ARef1}) that a system of relaxation
gasdynamics equations reduces to the equation
\begin{eqnarray}
\nonumber
\frac{\partial u}{\partial y}\pm \beta \frac{\partial u}{\partial x}
\pm \frac{\gamma M^3}{2\beta u_{s}}u\frac{\partial u}{\partial x}
=\pm \frac{\mu M^3}{u_{s}}\frac{\partial^2 u}{\partial x^2}
\mp\frac{M\alpha}{2\beta}u
\end{eqnarray}
with the boundary conditions
\begin{eqnarray}
\nonumber
u(y=\pm0)=\pm
\frac{Mu_s}{\beta}[\frac{\partial \xi_{2;1}}{\partial x} + \\
+ \mu \frac{M^3}{2\beta^2} \frac{\partial^2 \xi_{2;1}}{\partial x^2}
+ \frac{\alpha M}{2\beta^2}\xi_{2;1}].
\end{eqnarray}
Here $u\,=\,\partial \phi / \partial x$, $\phi$ is velocity disturbances
potential, $\beta\;=\;\sqrt {M^2-1}$, $\alpha\,=\,\alpha(\zeta) \,<\,0$
is an acoustic increment, $y\,=\,\xi_{2;1} (x)$ are equations of upper
and lower body surface.

Using equation (1) we obtain the coefficients of resistance $C_x$ and
lifting force $C_y$. In the limit case of homogeneous nonequilibrium
particles distribution the expressions for these coefficients is reduced
to form:
\begin{eqnarray}
\nonumber
C_x\,=\,\frac{2}{\beta}\left[2\delta^2 +<\theta_1^2> +
<\theta_2^2> + \frac{M\delta^2}{2 \beta} \alpha l_x \right],\\
\nonumber
C_y\,=\,\frac{2}{\beta} \left[2\delta + \frac{M\delta}{\beta^2}
\alpha l_x \right],
\end{eqnarray}
where $\theta_{2;1}(x)=\partial \xi_{2;1} (x)/\partial x\,-\,\delta$,
$\xi_{2;1} (0)\,=\,l_y\;,\;\xi_{2;1} (l_x)\,=\,0$.

Our main result is some decrease of $C_x$ and $C_y$ at
$\alpha \,<\,0$, i.e. in media with negative second viscosity.

The result, presented above, is only correct for laminated streams.
In the nonequilibrium media the bounds of laminar region is changed.
The traditional neglect of the second viscosity in subsonics is not correct if
$\zeta\,\gg\,\eta$, where $\eta$, is the shear viscosity coefficient.
For example, in $CO_2 , N_2 O$ - gases the relation $\zeta / \eta \,>\,10^3$,
as $T\,>\, 300 K$. (\cite{6131ARef2}). Therefore, the second viscosity is
essential for the sound propagation and the supersonic flows.

It is shown that such neglect is only correct under following conditions
\begin{eqnarray}
\nonumber
|\zeta| / \eta \ll 1 / M^2 , R \geq 1;\;
|\zeta| / \eta \ll R / M^2 , R \ll 1.
\end{eqnarray}
where R is Reynolds number. In other cases the second viscosity can even be
impotant for subsonic flows.

Taking into account these conditions the critical Reynolds number is obtained
\begin{eqnarray}
R_c \,=\,min|(D+B)/T|,
\end{eqnarray}
where
\begin{eqnarray}
\nonumber
D=\smallint v_{ik}^2 dV,
B=\frac{\zeta}{\eta} \smallint v_{ii}^2 dV,
T=\smallint v_i v_k v_{ik}dV,
\end{eqnarray}
$v_{ik}\,=\,\partial v_i / \partial x_k$, $\vec{v}$ is a disturbance of
the main flow.

At $\zeta\,=\,0$, condition (2) is similar to one from the (\cite{6131ARef1}).
At $\zeta\,>\,0$, the instability threshold increases comparatively with the (\cite{6131ARef1}).
Otherwise at $\zeta\,<\,0$, but $|B|\,<\,D$ the instability threshold decreases. Finally,
at $|B|\,>\,D$ the critical nomber does not exist. Moreover, at $R\,<\,R_c$ the flow
is unstable relatively to every low frequency disturbances.

The condition $|B|\,\geq\,D$ is satisfied with Mach numbers $M\,\geq\,(\eta / |\zeta|)^{1/4}$ if $R\,\geq\,1$ or $M\,\geq\,(R^2 \eta / |\zeta|)^{1/4}$ if $R\,\ll\,1$.

\end{document}